\documentclass{pasj00}

\begin{document}
\SetRunningHead{M.~Sato et al.}{Distance to NGC~281}
\Received{2008/02/29}
\Accepted{2008/06/04}

\title{Distance to NGC~281 in a Galactic Fragmenting Superbubble:\\ Parallax Measurements with VERA}
\author{
Mayumi \textsc{Sato},\altaffilmark{1,2}
Tomoya \textsc{Hirota},\altaffilmark{1,3}
Mareki \textsc{Honma},\altaffilmark{1,3}
Hideyuki \textsc{Kobayashi},\altaffilmark{1,2,3}
Tetsuo \textsc{Sasao},\altaffilmark{4,5}\\
Takeshi \textsc{Bushimata},\altaffilmark{1,6}
Yoon Kyung \textsc{Choi},\altaffilmark{1,2}
Hiroshi \textsc{Imai},\altaffilmark{7}
Kenzaburo \textsc{Iwadate},\altaffilmark{1}
Takaaki \textsc{Jike},\altaffilmark{1}\\
Seiji \textsc{Kameno},\altaffilmark{7}
Osamu \textsc{Kameya},\altaffilmark{1,3}
Ryuichi \textsc{Kamohara},\altaffilmark{1}
Yukitoshi \textsc{Kan-ya},\altaffilmark{8}
Noriyuki \textsc{Kawaguchi},\altaffilmark{1,3}\\
Mi Kyoung \textsc{Kim},\altaffilmark{1,2}
Seisuke \textsc{Kuji},\altaffilmark{1}
Tomoharu \textsc{Kurayama},\altaffilmark{5}
Seiji \textsc{Manabe},\altaffilmark{1,3}
Makoto \textsc{Matsui},\altaffilmark{9}\\
Naoko \textsc{Matsumoto},\altaffilmark{9}
Takeshi \textsc{Miyaji},\altaffilmark{1}
Takumi \textsc{Nagayama},\altaffilmark{9}
Akiharu \textsc{Nakagawa},\altaffilmark{7}
Kayoko \textsc{Nakamura},\altaffilmark{9}\\
Chung Sik \textsc{Oh},\altaffilmark{1,2}
Toshihiro \textsc{Omodaka},\altaffilmark{7}
Tomoaki \textsc{Oyama},\altaffilmark{1}
Satoshi \textsc{Sakai},\altaffilmark{1}
Katsuhisa \textsc{Sato},\altaffilmark{1}\\
Katsunori M. \textsc{Shibata},\altaffilmark{1,3}
Yoshiaki \textsc{Tamura},\altaffilmark{1,3}
and Kazuyoshi \textsc{Yamashita}\altaffilmark{3}
}

\altaffiltext{1}{Mizusawa VERA Observatory, National Astronomical Observatory, 2-12 Hoshi-ga-oka, Mizusawa-ku, Oshu, Iwate 023-0861}
\email{mayumi.sato@nao.ac.jp}
\altaffiltext{2}{Department of Astronomy, Graduate School of Science, The University of Tokyo, 7-3-1 Hongo, Bunkyo-ku, Tokyo 113-0033}
\altaffiltext{3}{Department of Astronomical Sciences, Graduate University for Advanced Studies, 2-21-1 Osawa, Mitaka, Tokyo 181-8588}
\altaffiltext{4}{Department of Space Survey and Information Technology, Ajou University, Suwon, Republic of Korea}
\altaffiltext{5}{Korean VLBI Network, KASI, Seoul, Republic of Korea}   
\altaffiltext{6}{Space VLBI Project, National Astronomical Observatory, 2-21-1 Osawa, Mitaka, Tokyo 181-8588}
\altaffiltext{7}{Faculty of Science, Kagoshima University, 1-21-35 Korimoto, Kagoshima, Kagoshima 890-0065}
\altaffiltext{8}{Department of Astronomy, Yonsei University, Seoul, Republic of Korea}
\altaffiltext{9}{Graduate School of Science and Engineering, Kagoshima University, 1-21-35 Korimoto, Kagoshima, Kagoshima 890-0065}
\maketitle

\begin{abstract}
We have used the Japanese VLBI array VERA to perform high-precision astrometry of an H$_2$O maser source in the Galactic star-forming region NGC~281~West, which has been considered to be part of a 300-pc superbubble.
We successfully detected a trigonometric parallax of 0.355$\pm$0.030~mas, corresponding to a source distance of 2.82$\pm$0.24~kpc.
Our direct distance determination of NGC~281 has resolved the large distance discrepancy between previous photometric and kinematic studies; likely NGC~281 is in the far side of the Perseus spiral arm.
The source distance as well as the absolute proper motions were used to demonstrate the 3D structure and expansion of the NGC~281 superbubble, $\sim$650~pc in size parallel to the Galactic disk and with a shape slightly elongated along the disk or spherical, but not vertically elongated, indicating the superbubble expansion may be confined to the disk. 
We estimate the expansion velocity of the superbubble as $\sim 20$~km~s$^{-1}$ both perpendicular to and parallel to the Galactic disk with a consistent timescale of $\sim 20$~Myr.
\end{abstract}

\section{Introduction}
A paradigm shift has been occurring in recent years in the observational study of the interstellar medium (ISM) in the Galaxy and in external galaxies, where the ISM is now recognized as being of fundamental importance not only in the life cycle of stars but also in galactic evolution. 
The present understanding of the ISM draws a dynamic and violent picture of the gas in the vast interstellar space, rather than a static one envisioned previously.
Recent observations with higher angular resolution have revealed the existence of dynamic and diverse gas structures in the Galaxy that cannot be explained by the traditional evolutionary cycle of the ISM, requiring new insights toward a comprehensive understanding of the ISM (e.g., Reynolds 1997, 2002; Dickey 2001).
Extreme activity of the ISM is indicated by large-scale circular or arc-like structures (up to a few kiloparsecs in scale) called superbubbles or supershells, many of which were identified earlier by Heiles (1979, 1984) based on H\emissiontype{I} 21-cm line surveys by Weaver and Williams (1973) and Heiles and Habing (1974).
Kiss, Mo\'or, and T\'oth (2004) later identified 145 large-scale loop- or arc-like structures in the 2nd Galactic quadrant (i.e.\ for galactic longitudes $90^\circ < l < 180^\circ$) on {\it IRAS}-based far-infrared maps.
Recent results for the 1st, 3rd, and 4th Galactic quadrants by K\"onyves \etal\ (2007) increased the number of identified loops to 462.
Similar objects are also found in nearby galaxies, including the Local Group spirals Andromeda (M31; Brinks \& Bajaja 1986) and M33 (Deul \& den Hartog 1990), Local Group irregulars including the Large Magellanic Cloud (LMC; Chu \etal\ 1993, 1995; Kim \etal\ 1998, 1999), the Small Magellanic Cloud (SMC; Staveley-Smith \etal\ 1997; Stanimirovi\'c \etal\ 1999) and IC~10 (Shostak \& Skillman 1989), large spiral galaxies such as M101 (Kamphuis \etal\ 1991) and NGC~891 (Howk \& Savage 1997, 2000), and dwarf galaxies such as M82 (Matsushita \etal\ 2000; Matsumoto \etal\ 2001), IC~2574, DDO~47 (Walter \& Brinks 1999, 2001) and Holmberg~II (HoII; Puche \etal\ 1992).
\begin{figure*}[t]
  \begin{center}
    \FigureFile(86mm,110mm){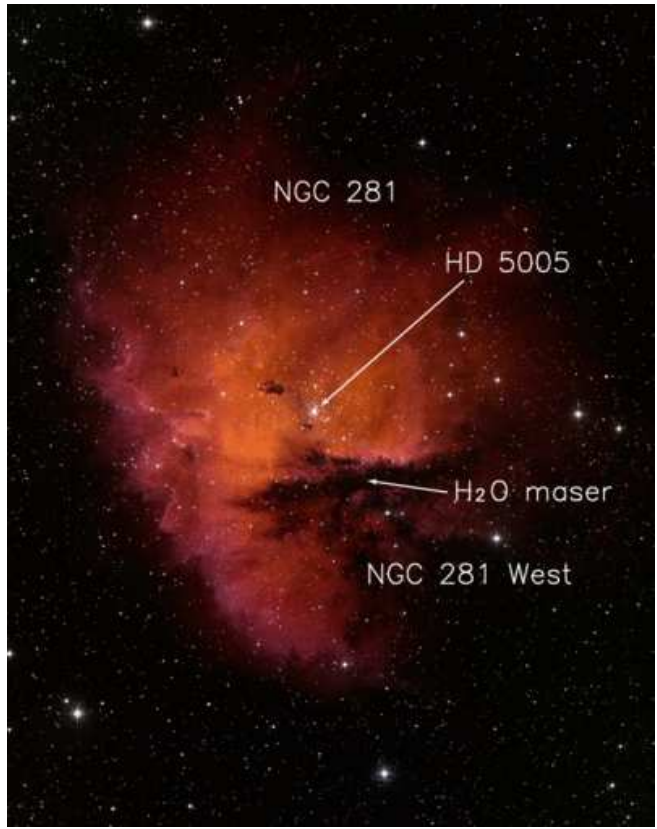}
  \end{center}
  \caption{A wide-field optical image of the star-forming region NGC~281 taken with the WIYN 0.9-meter at Kitt Peak National Observatory (courtesy of T.A. Rector/University of Alaska Anchorage and WIYN/AURA/NSF). The orientation of the image is north up and east left. The red nebulosity of the H\emissiontype{II} region NGC~281 is obscured in its southwestern quadrant by the adjoining molecular cloud NGC~281~West, where the H$_2$O maser source is located.}\label{fig:optical}
\end{figure*}

The most widely accepted theory for the origin of superbubbles suggests that supernova explosions and strong stellar winds from OB associations in a galactic disk blow the surrounding gas up into the halo, carving out extensive cavities in the ISM (see Tenorio-Tagle \& Bodenheimer 1988).
Superbubbles are therefore believed to play a vital role in disk-halo interactions, serving as a conduit for matter and energy from the disk into the halo in the form of galactic {`}fountains{'} (Shapiro \& Field 1976) or {`}chimneys{'} (Norman \& Ikeuchi 1989).

In the Galaxy, only a small number of chimneys (or also called {`}worms{'}) or fragmenting superbubbles/supershells (on several hundred to one thousand parsec scales) are known and well-studied: e.g., the Orion-Eridanus superbubble (Cowie \etal\ 1979; Reynolds \& Ogden 1979; Boumis \etal\ 2001; Welsh \etal\ 2005), the Cygnus superbubble (Cash \etal\ 1980), the Stockert chimney (M\"uller \etal\ 1987; Kundt \& M\"uller 1987; Forbes 2000), the Aquila supershell (Maciejewski \etal\ 1996), the Scutum supershell (Callaway \etal\ 2000; Savage \etal\ 2001), GSH~277+00+36 (McClure-Griffiths \etal\ 2000), GSH~242-03+37 (McClure-Griffiths \etal\ 2006), the W4 chimney/superbubble (Normandeau \etal\ 1996; Reynolds \etal\ 2001; Terebey \etal\ 2003; Madsen \etal\ 2006; West \etal\ 2007), the Ophiuchus superbubble (Pidopryhora \etal\ 2007), and the NGC~281 superbubble (Megeath \etal\ 2002, 2003; Sato \etal\ 2007).
Superbubbles are also a key target of research for understanding their impact on the history of star formation in the galactic disk, including but not limited to superbubble-triggered star formation (e.g., NGC~281 as explained later in the section).

In either the Galaxy or external galaxies, very little is known about the dynamics and energetics of superbubbles/chimneys or about their influence on the formation of the halo or on the star formation history in the galactic disk.
A difficulty in revealing the structure and kinematics of superbubbles stems from the limited information obtainable from two dimensions of sky positions and velocity information only along the line of sight.
A direct observation of the three-dimensional (3D) motion and structure of a superbubble has not been easy due to the high astrometric accuracy required.
However, Sato \etal\ (2007) reported the first VLBI (Very Long Baseline Interferometry) observations that obtained direct evidence for expanding motions of a superbubble by measuring the proper motions.

In Sato \etal\ (2007), we observed the H$_2$O maser emission in NGC~281~West over 6 months with VERA (VLBI Exploration of Radio Astrometry), a new Japanese radio telescope array dedicated to phase-referencing VLBI astrometry (e.g., Honma \etal\ 2000, 2005, 2007; Kobayashi \etal\ 2003; Imai \etal\ 2006, 2007; Hirota \etal\ 2007, 2008a, 2008b; Nakagawa \etal\ 2008; Kim \etal\ 2008; Choi \etal\ 2008).
These observations revealed the systemic motion of the NGC~281 superbubble away from the Galactic plane at a velocity of 20$-$30~km~s$^{-1}$.
At its estimated heliocentric distance of 2.2$-$3.5~kpc, NGC~281 most likely originated in the Galactic plane and has been blown out by a superbubble expansion (see $\S$1.1 and Sato \etal\ 2007).

A precise and direct determination of the distance to NGC~281 is desirable and essential for a better understanding of the region, including further description of the 3D structure and motion of the superbubble to reveal the origin, the energetics and the timescale of the superbubble.
These are the primary reasons why we decided to complete our VERA observations to directly measure the parallactic distance to NGC~281 via an H$_2$O maser source.
The astrometric determination of the distance to NGC~281 is also important for studying early-type (O$-$B) high-mass ($M\gtrsim10\Mo$) stars and stellar evolution in combination and in comparison with photometric studies of the OB stars in NGC~281 (e.g., Henning \etal\ 1994; Guetter \& Turner 1997). 
In this paper, we present the results of our parallax measurements with VERA over 18 months, which enables us to further discuss the 3D structure, expanding motion, the origin and the timescale of the NGC~281 superbubble.

\subsection{NGC~281 Superbubble}
The NGC~281 region ($\alpha_{2000}=$00$^{\rm h}$52$^{\rm m}$, $\delta_{2000}=+$56$^\circ$\timeform{34'} or $l=$123$^\circ$.07, $b=-$6.$^\circ$31) provides an excellent laboratory for studying in detail the cycle of the ISM and its impact on both star formation and Galactic disk-halo interaction through superbubbles.
At its estimated heliocentric distance of 3~kpc, this region is remarkably located far above ($\sim 300$~pc) the midplane of the Perseus arm of the Galaxy.
Of special interest in the NGC~281 region is a possibility of triggered star formation occurring on two different scales: large-scale ($\sim$300~pc) supperbubble-triggered formation of the first OB stars (Megeath \etal\ 2002, 2003), and the sequential and ongoing triggered star formation ($\sim$1$-$10~pc) in an adjoining giant molecular cloud (NGC~281~West) through interaction with an H\emissiontype{II} region (the NGC~281 nebula) excited by the first OB stars (Elmegreen \& Lada 1978; Megeath \& Wilson 1997).

Figure~\ref{fig:optical} is an optical image of the region, which clearly shows the red nebulosity of the H\emissiontype{II} region NGC~281 (also known as Sharpless 184) of diameter 20~arcmin ($\sim 17$~pc), in which an early-type cluster called IC~1590 is embedded (Guetter \& Turner 1997).
The brightest member of the cluster IC~1590 is an O-type Trapezium-like system HD~5005 (or also called ADS~719), whose component stars HD~5005ab (unresolved), HD~5005c, and HD~5005d have spectral types of O6.5~V, O8~V, and O9~V, respectively (Walborn 1973; Abt 1986; Guetter \& Turner 1997).

The southwestern quadrant of the NGC~281 nebula is obscured (as seen in figure~\ref{fig:optical}) by the adjoining molecular cloud NGC~281~West in front of the nebula.
Ongoing star formation in the NGC~281~West cloud is indicated by the H$_2$O maser emission and {\it IRAS} sources within this cloud near the clumpy interface between the H\emissiontype{II} region and the cloud.
This generation of stars may have been triggered by interactions with the H\emissiontype{II} region (Elmegreen \& Lada 1978; Megeath \& Wilson 1997).
The NGC~281 molecular cloud complex that surrounds the H\emissiontype{II} region (including the NGC~281~West and East clouds) was mapped in CO emission lines by Lee and Jung (2003).
The central radial velocity of the NGC~281~West molecular cloud, $V_{\rm{LSR}}=31$~km~s$^{-1}$ (Lee and Jung 2003), agrees well with that of the H$_2$O maser emission in the cloud (Sato \etal\ 2007).  
On a larger scale, Megeath \etal\ (2002, 2003) identified the CO molecular cloud complex as being on an H\emissiontype{I} loop extending over 300~pc away from the Galactic plane toward decreasing galactic latitude.
In an $l$ vs.\ $V_{\rm{LSR}}$ diagram (i.e.\ a plot of galactic longitude vs.\ observed radial velocity), the CO molecular clouds appear to be part of a broken ring of diameter 270~pc, expanding at a velocity of 22~km~s$^{-1}$ parallel to the Galactic plane (Megeath \etal\ 2002, 2003).
Megeath \etal\ (2002, 2003) suggested that these clouds were formed in a fragmenting superbubble, which triggered the formation of the first OB stars, and that these OB stars then ionized the surrounding gas which subsequently triggered ongoing star formation in the neighboring cloud.

In Sato \etal\ (2007), we derived the absolute proper motions of the H$_2$O maser features in NGC~281~West by phase-referencing VLBI observations with VERA over 6 months.
We found 10 maser features, highly likely excited by two spatially distinct young stellar objects (YSOs), are systematically moving toward southwest, in agreement with the direction of the proper motion of HD~5005 in the H\emissiontype{II} region measured with larger uncertainties by {\it Hipparcos} (Perryman \etal\ 1995, 1997).
Our results with VERA revealed the systemic motion of the NGC~281 region away from the Galactic plane, most likely having originated in the Galactic plane and been blown out by the expansion of the NGC~281 fragmenting superbubble.

The uncertainties in our velocity estimates are largely due to the large distance uncertainty of NGC~281, despite many measurements, e.g., photometric distances of 2~kpc (Sharpless 1952, 1954), 3.68~kpc (Cruz-Gonz\'alez \etal\ 1974), 3.5~kpc (Henning \etal\ 1994) and 2.94$\pm$0.15~kpc (Guetter \& Turner 1997) and kinematic distances of 2.2~kpc (Georgelin \& Georgelin 1976; Roger \& Pedlar 1981) and 3~kpc (Lee \& Jung 2003).
In Sato \etal\ (2007), we adopted a value of 2.9~kpc as the distance to NGC~281, based on a recent estimate of 2.94$\pm$0.15~kpc derived by Guetter and Turner (1997) through photometry of 279 individual stars in and about the young cluster IC~1590 embedded in the NGC~281 nebula.

The most reliable and direct way to determine the distance ($d$) of an astronomical object is to measure the trigonometric parallax ($\pi$) of the object due to the Earth's motion around the solar system barycenter, where the distance is given by $d$(pc) $=1/\pi$(\timeform{"}). 
The astrometric determination of the distance to NGC~281 provides valuable information for studying the formation and evolution of early-type (O$-$B) high-mass ($M\gtrsim10\Mo$) stars in comparison with photometric studies and distances, for example, by Guetter \& Turner (1997).

In the present study, we report on our successful determination of the parallactic distance to NGC~281, which reveals the 3D structure of the NGC~281 fragmenting superbubble, and investigate the origin and timescale of the formation of such a large-scale ($\sim$650~pc) structure.

\section{VERA Observations and Data Reduction}
The VERA observations were carried out over 13 epochs spaced approximately at monthly to bimonthly intervals.
The first 6 epochs reported in Sato \etal\ (2007) were on 2006 May 14, July 21, August 3, September 5, October 25, and November 18 (days of year 134, 202, 215, 248, 298, and 322, respectively).
Seven new observations were conducted on 2006 December 11, 2007 January 22, March 19, May 1, July 27, September 30, and October 28 (days of year 345, 387, 443, 486, 573, 638, 666 as counted from 2006 January 0).
Note that the dates stated here are universal time (UT) start times, while for the data analyses in the following section we use more precisely the median hour of each observation, which lasted 7$-$9 hours.

The observational procedures are as described in Sato \etal\ (2007) and as follows. 
We observed two sources simultaneously in the dual-beam mode of VERA for phase referencing (e.g., Honma \etal\ 2003), and the real-time instrumental phase difference data between the two beams were taken with an artificial noise source in each beam and recorded for calibration (Kawaguchi \etal\ 2000; Honma \etal\ 2008a). 
We simultaneously observed the H$_2$O maser source in the NGC~281~West cloud with one of the two extragalactic position-reference quasars J0047+5657 (0$^\circ$.84 separation; at $\alpha_{2000}=$00$^{\rm h}$47$^{\rm m}$00.$^{\rm s}$428805, $\delta_{2000}=+$56$^\circ$\timeform{57'}\timeform{42.''39479}) and J0042+5708 (1$^\circ$.50 separation; at $\alpha_{2000}=$00$^{\rm h}$42$^{\rm m}$19.$^{\rm s}$451727, $\delta_{2000}=+$57$^\circ$\timeform{08'}\timeform{36.''58602}).
Positions are from Beasley \etal\ (2002).
We alternated between the two quasars typically every 10 minutes.
We observed a bright calibrator source hourly: J0319+4130 ($=$ 3C~84) at the first epoch (2006 May 14) and J2232+1143 ($=$ CTA~102) for the other 12 epochs.
The data correlation was performed with the Mitaka FX correlator (Chikada \etal\ 1991) with frequency and velocity resolutions for the maser lines of 15.625~kHz and 0.21~km~s$^{-1}$, respectively.
The observed frequencies of the maser lines were converted to radial (line-of-sight) velocities with respect to the local standard of rest (LSR), $V_{\rm LSR}$, using a rest frequency of 22.235080~GHz for the H$_2$O 6$_{16}$-5$_{23}$ transition.

Visibility calibration and imaging were performed in a standard manner with the NRAO Astronomical Image Processing System (AIPS) package.
The visibilities of all H$_2$O maser channels at each epoch were phase-referenced to each of the position-reference sources J0047+5657 and J0042+5708 independently by applying the phase solutions from fringe fitting (AIPS taks FRING) for each position-reference source to the H$_2$O maser channels for the corresponding observation time and frequencies.
The instrumental phase difference between the two beams was also corrected using the recorded phase-calibration data described above.
We also calibrated the error in the visibility phase caused by the Earth's atmosphere based on GPS measurements of the atmospheric zenith delay due to tropospheric water vapor (Honma \etal 2008b). 
After calibration, we made spectral-line image cubes using the AIPS task IMAGR around the maser spots with 512 pixels $\times$ 512 pixels of size 0.05 mas.
The synthesized beam had an FWHM beam size of 1.3~mas $\times$ 0.8~mas with a position angle of $-$43$^\circ$ East of North.
Least-squares fittings were performed to simultaneously solve for the sinusoidal parallax curve and linear proper motion in the directions of right ascension (RA) and declination (Dec).

\section{Results}
\begin{figure*}[htbp]
  \begin{center}
    \FigureFile(150mm,300mm){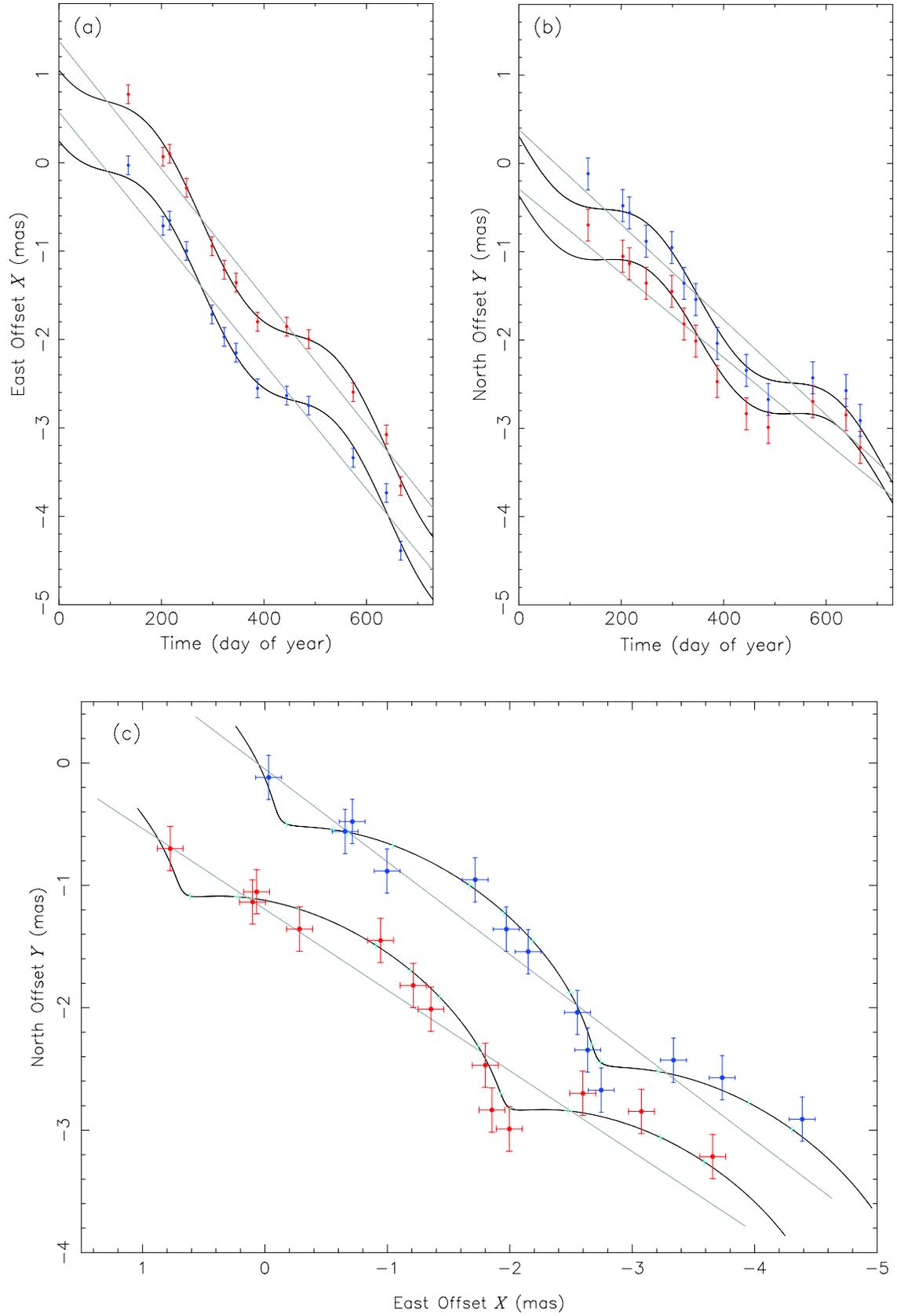}
  \end{center}
  \caption{Position measurements of the peak maser spot of feature~4; using J0047+5657 (blue) and J0042+5708 (red) as position-reference sources. The coordinate origin is at $\alpha_{2000}=$00$^{\rm h}$52$^{\rm m}$24.$^{\rm s}$70081, $\delta_{2000}=+$56$^\circ$\timeform{33'}\timeform{50.''5274}. (a) and (b) show the eastward ($X=$RA~cos(Dec)) and northward ($Y=$Dec) positional offsets in mas, versus time (day of year, as counted from 2006 January 0), respectively. Best-fit models are also plotted with gray lines for proper motions and black curves for the sum of sinusoidal parallax and the linear proper motions.  (c) The trajectory of the positional offsets $X$ versus $Y$ (in mas) on the sky.  Best-fit models are also plotted, with cyan crosses showing the expected positions on each day of observation.  The associated error bars in each panel indicate the standard deviation in $X$ or $Y$ from the least-squares fits (see text).}\label{fig:parallax}
\end{figure*}

\subsection{Parallax Measurements}
\begin{table*}
 \begin{center}
 \caption{Fitted distances and absolute proper motions for the peak maser spot of feature~4.}
  \begin{tabular}{lcccccc}
   \hline
   \multicolumn{1}{c}{$X$ , $Y$}& {Reference} & $\pi$ [mas] & $d$ [kpc]& $\sigma$ [mas] & $\mu_X$ [mas~yr$^{-1}$]& $\mu_Y$ [mas~yr$^{-1}$]\\
   \multicolumn{1}{c}{(1)}&(2)&(3)&(4)&(5)&(6)&(7)\\
   \hline\hline
   {$X$}  & J0047+5657                  & 0.346~(0.052)  & 2.89~(0.43) & 0.113 & $-2.60$~(0.07) & \\
   {(RA)} & J0042+5708                  & 0.334~(0.047)  & 2.99~(0.42) & 0.104 & $-2.65$~(0.07) & \\
   \cline{2-7}
   ~~     & Combined                    & 0.340~(0.034)  & 2.94~(0.30) & 0.106 & $-2.63$~(0.05) & \\
   \hline
   {$Y$}  & J0047+5657                  & 0.400~(0.095)  & 2.50~(0.59) & 0.188 &         & $-1.97$~(0.12)\\
   {(Dec)}& J0042+5708                  & 0.412~(0.092)  & 2.43~(0.54) & 0.183 &         & $-1.75$~(0.11)\\
   \cline{2-7}
   ~~     & Combined                    & 0.406~(0.065)  & 2.46~(0.39) & 0.181 &         & $-1.86$~(0.08)\\
   \hline
   {$X$ \& $Y$}& Combined               & 0.355~(0.030)  & 2.82~(0.24) & 0.148 & &\\  
   \hline
   \multicolumn{7}{@{}l@{}}{\hbox to 0pt{\parbox{150mm}{\footnotesize
(1) Directions of position measurement: eastward ($X$) and northward ($Y$).  Data were uniformly weighted for each epoch.  For the X \& Y direction, data in the two directions $X$ and $Y$ were error-weighted with the standard deviation in the fitting in each direction.  
(2) Position-reference sources used for the measurement. {`}Combined{'} indicates the simultaneous fitting of two independent position-reference results using J0047+5657 and J0042+5708.  
(3) Measured parallaxes (in mas) with associated uncertainties in parentheses.
(4) Distances corresponding to the parallax (in kpc) with associated uncertainties in parentheses.
(5) The standard deviation of the post-fit residuals (in mas) from the least-squares fit.
(6)(7) Absolute proper motions in $X$ and $Y$ (in mas~yr$^{-1}$) with associated uncertainties in parentheses.
     }\hss}}
   \end{tabular}
 \end{center}
 \end{table*}

Figure~\ref{fig:parallax} shows the position measurements of the brightest maser spot at $V_{\rm{LSR}}=-32.1$~km~s$^{-1}$ (feature \#4 of 10 features identified in Sato \etal\ 2007) using J0047+5657 (blue) and J0042+5708 (red) as position-reference sources independently.
Figures~\ref{fig:parallax}a and \ref{fig:parallax}b show the positional offsets of the maser spot in the eastward ($X$, i.e.\ (RA)~cos(Dec)) and northward ($Y$, i.e.\ Dec) directions, respectively, relative to the reference position at the origin at $\alpha_{2000}=$00$^{\rm h}$52$^{\rm m}$24.$^{\rm s}$70081 for $X$ and $\delta_{2000}=+$56$^\circ$\timeform{33'}\timeform{50.''5274} for $Y$, as a function of time (day of year, as counted from 2006 January 0).
The constant offsets of $\sim 0.80$~mas seen in each of $X$ and $Y$ between the two measurements using different position-reference sources J0047+5657 and J0042+5708 are due to the uncertainties in the absolute positions of these reference sources (0.64~mas for J0047+5657 and 0.82~mas for J0042+5708; Beasley \etal\ 2002) and are solved for in the multi-epoch parallax and proper-motion fittings. 
The best-fit models are also plotted with gray lines for proper motions and black curves for the sum of sinusoidal parallax and the linear proper motions.
Table~1 summarizes the fitted distances and proper motions.
Figures~\ref{fig:parallax}c is a plot of the spot position in $Y$ vs.\ $X$ and its movement during the observation epochs.

The obtained values of the parallax in $X$ are: $\pi=0.346\pm0.052$~mas (corresponding to a distance of $d=2.89\pm0.43$~kpc) using J0047+5657 and $\pi=0.334\pm0.047$~mas (corresponding to a distance of $d=2.99\pm0.42$~kpc) using J0042+5708.
Combining these independent measurements using two different position-reference sources, we have obtained as a final value in $X$ (RA) direction: $\pi=0.340\pm0.034$~mas ($d=2.94\pm0.30$~kpc).

Similarly, we have obtained the parallax in $Y$ to be: $\pi=0.400\pm0.095$~mas ($d=2.50\pm0.59$~kpc) using J0047+5657 and $\pi=0.412\pm0.092$~mas ($d=2.43\pm0.54$~kpc) using J0042+5708.
Combining these two measurements yields the final value in $Y$ (Dec) direction to be: $\pi=0.406\pm0.065$~mas ($d=2.46\pm0.39$~kpc).
Here we have estimated the associated uncertainties from the uniformly-weighted standard deviation $\sigma$ from the least-squares fitting in $X$ and $Y$, which were $\sigma_X=0.104-0.113$~mas and $\sigma_Y=0.181-0.188$~mas, respectively (see table~1 for the individual value of the standard deviation in each measurement). 
Random errors in individual position measurements estimated with the AIPS Gaussian-fit task JMFIT caused by the noise (which are approximated by the beam HWHM divided by the signal-to-noise ratio) were $\sim 0.010$~mas, which is thus not the main cause of the deviations $\sigma_X$ and $\sigma_Y$ from the fits.
We will discuss the possible error sources in $\S4.1$ in more detail.
However, those errors are difficult to measure quantitatively, and we therefore estimated the astrometric errors by the standard deviations of the residuals from the fits and plotted the standard deviations as the error bars of each point in figure~\ref{fig:parallax}.

Weighting by $\sigma_X$ and $\sigma_Y$, we simultaneously fitted the position measurements of the maser spot in both $X$ and $Y$ altogether, including all data using J0047+5657 and J0042+5708, to obtain the final result for the parallax measurements to be: $\pi=0.355\pm 0.030$~mas, which corresponds to a distance to NGC~281 of $d=2.82\pm 0.24$~kpc.
We have therefore successfully determined the parallactic distance to NGC~281 as far as $\sim3$~kpc with a precision better than 10\%.
Our distance estimate of NGC~281, $d=2.82\pm 0.24$~kpc, agrees well with the recent photometric distance of 2.94$\pm$0.15~kpc by Guetter and Turner (1997), but excludes the kinematic distance of 2.2~kpc by Georgelin and Georgelin (1976) and by Roger and Pedlar (1981) and most distance estimates in earlier studies.

\subsection{Absolute Proper Motions}
\begin{table*}
 \begin{center}
 \caption{Comparison of the absolute proper motions of feature~4 between the present study and Sato \etal\ (2007).}
  \begin{tabular}{ccccc}
   \hline
   \multicolumn{1}{c}{Reference} & $\mu_{X}$ [mas~yr$^{-1}$] & $\mu_{X}$ [mas~yr$^{-1}$]& $\mu_{Y}$ [mas~yr$^{-1}$] & $\mu_{Y}$ [mas~yr$^{-1}$]\\
   \multicolumn{1}{c}{}& Present study & Sato \etal\ (2007)& Present study & Sato \etal\ (2007)\\
   \multicolumn{1}{c}{(1)}&(2)&(3)&(4)&(5)\\
   \hline\hline
   J0047+5657    & $-2.60$~(0.07) & $-2.87$~(0.26) & $-1.97$~(0.12) & $-2.78$~(0.37)\\
   J0042+5708    & $-2.65$~(0.07) & $-2.92$~(0.23) & $-1.75$~(0.11) & $-2.54$~(0.35)\\
   \hline
   Combined      & $-2.63$~(0.05) & $-2.89$~(0.18) & $-1.86$~(0.08) & $-2.65$~(0.26)\\
   \hline
  \end{tabular}\\
{\parbox{150mm}{\footnotesize
(1) Position-reference sources used for the measurement. {`}Combined{'} indicates the simultaneous fitting of two independent position-reference results using J0047+5657 and J0042+5708.  
(2)(3) Absolute proper motions of feature~4 derived in the present study and in Sato \etal\ (2007), respectively, in $X$ direction (eastward) (in mas~yr$^{-1}$) with associated uncertainties in parentheses.
(4)(5) Same as (2)(3) but in $Y$ direction (northward).
     }\hss}
 \end{center}
\end{table*}

As listed in table~1, we fitted the position measurements of 13 epochs simultaneously for both the parallax $\pi$ and proper motions ($\mu_X$,~$\mu_Y$) (in RA and Dec, respectively) of the peak maser spot of feature~4 (the brightest feature in the maser component C3, which is one of the two spatial components C1 and C3 with maser emission detected in Sato \etal\ 2007).
For the proper motions, we obtained ($-$2.60$\pm$0.07~$-$1.97$\pm$0.11) mas~yr$^{-1}$ using J0047+5657, and ($-$2.65$\pm$0.07,~$-$1.74$\pm$0.11) mas~yr$^{-1}$ using J0042+5708.
The error-weighted mean proper motions of this spot (which are the same as unweighted mean in this case of equal errors) are: ($\mu_X$,~$\mu_Y$)$=$($-$2.63$\pm$0.05,~$-$1.86$\pm$0.08) mas~yr$^{-1}$.
We consider this mean as revised values (called {`}modified{'} hereafter) of the error-weighted mean proper motions (denoted by 4w) of feature~4 reported in Sato \etal\ (2007).

Table~2 shows a comparison of the absolute proper motions derived for feature~4 between the revised values in the present study and the previous values in Sato \etal\ (2007).
In Sato \etal\ (2007), we derived absolute proper motions of feature~4 (in C3) and features~9 and 10 (in C1, the other one of the two spatial maser components; see Sato \etal\ 2007) assuming the distance to NGC~281 to be $d=2.9$~kpc (adopting the photometric distance by Guetter and Turner 1997) and subtracting the expected parallax in the position measurements during the first 6 observation epochs.

The revised values of $\mu_X$ for feature~4 in the present study agree with the previous values in Sato \etal\ (2007) within the margin of errors ($1\sigma$), however, for $\mu_Y$ we find deviations of the revised values from the previous ones larger than the errors ($>2\sigma$) estimated in Sato \etal\ (2007).
This means in Sato \etal\ (2007) we overestimated $\mu_Y$ of feature~4 and underestimated the errors in the linear least-squares fittings (after subtracting the assumed parallax).
As seen in figure~\ref{fig:parallax} based on the present longer-time observations, the overestimate of the previous $\mu_Y$ values for feature~4 is due to a large position deviation from the best-fit at the first ($+2\sigma$) epochs, which yielded a larger slope.
The underestimate of the errors occured mostly due to non-gaussian errors larger than 0.1~mas (larger than thermal errors by an order of magnitude) in the position measurements, owing to astrometric error sources that are discussed in more detail in $\S4.1$.
However, as will be described in $\S4.2$, the scientific conclusions remain essentially unchanged after the modifications to the proper motions.

By adding the mean relative motion of all 8 features (including feature~4) in C3 with respect to the reference feature~4, ($\bar{\mu_x}$, $\bar{\mu_y}$)$=$(0.46$\pm$0.33,~0.42$\pm$0.31) mas~yr$^{-1}$ as given in Sato \etal\ (2007), to this modified absolute motion of feature~4, the mean absolute proper motion of these features in C3 (denoted by C3m in Sato \etal\ 2007) is then also revised to be: ($\bar{\mu_X}$,~$\bar{\mu_Y}$)$_{\rm C3}=$($-$2.17$\pm$0.33,~$-$1.44$\pm$0.32) mas~yr$^{-1}$.

Our new distance estimate of NGC~281 by parallax measurements, $d=2.82$~kpc, also introduces a slight modification to the measured absolute proper motions of features~9 and 10 on the same data of 6 epochs (as these features were not persistent through the total 13-epoch observations with high enough signal-to-noise ratios for the parallax and proper motion fittings).
The modified absolute proper motions ($\mu_X$,~$\mu_Y$) of feature~9 are ($-$4.85$\pm$0.11,~$-$1.73$\pm$0.46) mas~yr$^{-1}$ using J0047+5657, and ($-$4.74$\pm$0.28,~$-$1.44$\pm$0.41) mas~yr$^{-1}$ using J0042+5708.
Similarly we obtained the absolute proper motion ($\mu_X$,~$\mu_Y$) of feature~10 to be: ($-$2.00$\pm$0.23,~$-$4.06$\pm$0.79) mas~yr$^{-1}$ using J0047+5657, and ($-$1.89$\pm$0.38,~$-$4.14$\pm$0.69) mas~yr$^{-1}$ using J0042+5708.
Here the associated uncertainties in $\mu_X$ and $\mu_Y$ were estimated from the standard deviations from the linear least-squares fits in $X$ and $Y$ positional offsets (after the parallax for $d=2.82$~kpc is subtracted), respectively.
The error-weighted mean proper motions of each of these features in C1 (denoted by 9w and 10w in Sato \etal\ 2007) are ($\mu_X$,~$\mu_Y$)$=$($-$4.84$\pm$0.10,~$-$1.57$\pm$0.31) mas~yr$^{-1}$ for feature~9 and ($\mu_X$,~$\mu_Y$)$=$($-$1.97$\pm$0.20,~$-$4.11$\pm$0.52) mas~yr$^{-1}$ for feature~10.
Then the unweighted mean of the absolute motions of these two features in C1 (denoted by C1m in Sato \etal\ 2007) is also slightly modified to be ($\bar{\mu_X}$,~$\bar{\mu_Y}$)$_{\rm C1}=$($-$3.41,~$-$2.84) mas~yr$^{-1}$.

\section{Discussion}

\subsection{Astrometric Error Sources}
In the present work, we measured the annual parallax of NGC~281~West to be $0.355\pm 0.030$~mas, which corresponds to a distance of $2.82\pm 0.24$~kpc.
In this section, we first discuss possible error sources in our parallax and proper-motion measurements.
 
As already mentioned in $\S3.1$, the true uncertainties in the measurements were estimated from standard deviations of the post-fit residual from the least-squares fits (as listed in table~1), because thermal errors due to noise in individual position measurements ($\sim 0.010$~mas) are far smaller than the deviations from the fits.
The likely sources for such deviations are difficult to measure quantitatively, which led to our error underestimate in the derived absolute proper motions of feature~4 in Sato \etal\ (2007) as described in $\S3.2$.
The reference sources did not show any resolved structure such as additional components or structural change, and we obtained similar results.
Therefore we do not consider the contribution of the reference source as a dominant error source.

One of the likely causes for position uncertainties of $\sigma_X=0.104-0.113$~mas and $\sigma_Y=0.181-0.188$~mas (see table~1) is mis-modeling of tropospheric zenith delay, as also reported in previous astrometric measurements with VERA (e.g., Hirota \etal\ 2007; Nakagawa \etal\ 2008; Honma \etal\ 2008b).
The water vapor in the troposphere introduces optical path differences through the atmosphere between the target maser source and reference sources because of the elevation angle difference between the sources.
Therefore, even after the phase referencing calibration, there remain residuals of tropospheric zenith delay due to the water vapor, which are difficult to measure precisely and are one of the most serious error sources in the VLBI astrometry in the 22-GHz band.
Nakagawa \etal\ (2008) and Honma, Tamura \& Reid (2008b) give detailed discussion on astrometric errors caused by such residuals of tropospheric zenith delay calibration with VERA and report probable errors, $\sigma_X$ and $\sigma_Y$, of 0.05-mas and 0.10-mas levels, respectively, for source declination of 60$^\circ$.
Our parallax determination with VERA is no exception in having astrometric errors larger in declination ($Y$) than in right ascension ($X$) (i.e.\ $\sigma_Y>\sigma_X$) as the same tendencies reported in other observations with VERA (e.g., Hirota \etal\ 2007).
Larger errors in $Y$ (Dec) than in $X$ (RA) are explained by the tendency of tropospheric zenith delay errors to have more severe effects in the $Y$ direction (although depending of the separation angle between the target and reference sources and the declination of the sources) and are as expected by the simulation in Honma, Tamura and Reid (2008b).

However, we do not consider the residuals of tropospheric zenith delay as the dominant cause of errors in our parallax measurements. 
As the simulations by Honma, Tamura and Reid (2008b) clearly show, the astrometric errors due to the zenith delay residuals for the NGC~281 case at its high declination ($\delta_{2000}\simeq+$56$^\circ$.5) and with separation angle (SA) smaller than 1$^\circ$ for J0047+5206 (SA=0$^\circ$.84) are expected to be as good as $20-30$~$\mu$as (i.e.\ $0.020-0.030$~mas) in each of $X$ and $Y$ directions, which do not account for all the standard deviations $\sigma_X=0.104-0.113$~mas and $\sigma_Y=0.181-0.188$~mas in our measurements.
Also, as figure~\ref{fig:parallax} clearly demonstrates, using two different reference quasars with different separation angles (SA=0$^\circ$.84 for J0047+5206 and SA=1$^\circ$.50 for J0042+5708) did not result in remarkable differences in the astrometric errors for the two cases. 
This certainly excludes the simple zenith delay error effect as the main source of our astrometric errors.

The most likely major source of errors in the present position measurements is a variation of the intrinsic maser structure.
Most maser spots are unresolved in VLBI observations, however evidence has been shown for intrinsic maser structure smaller than 1~AU on submilliarcsecond scales in previous maser studies (e.g., Fish \& Sjouwerman 2007).
A maser spot likely consists of both unresolved substructure and observed larger structure which are indistinguishable (e.g., Fish \& Sjouwerman 2007), and thus a variation in maser substructure can lead to fluctuation of the peak position, limiting the accuracy in position measurements of a maser spot.
This effect of the maser spatial structure is also seen for other observations in parallax measurements with up to 10\% errors, independently of the source distance, for example, in Orion KL by Hirota \etal\ (2007). 
Errors due to the maser structure can be reduced by using multiple maser features for derivation of a parallax.
In our case of NGC~281~West, however, only one bright maser feature (above 10~Jy~beam$^{-1}$) that was used in the present study was found with a good signal-to-noise ratio for the precise parallax measurements, even persistent for the all observation epochs.

According to discussions by Honma \etal\ (2007), other possible astrometric errors in VERA observations that arise from the uncertainties in the antenna-station positions, delay models, and path length differences due to ionosphere are smaller by at least an order of magnitude than the uncertainties due to tropospheric zenith delay residuals. 
Therefore, we conclude that the main error sources of our astrometric measurements is the variability of maser feature structure, and the contribution by tropospheric zenith delay residuals is also nonnegligible.

\subsection{NGC~281 Superbubble vs.\ Galactic Rotation}
\begin{figure*}[t]
  \begin{center}
    \FigureFile(140mm,80mm){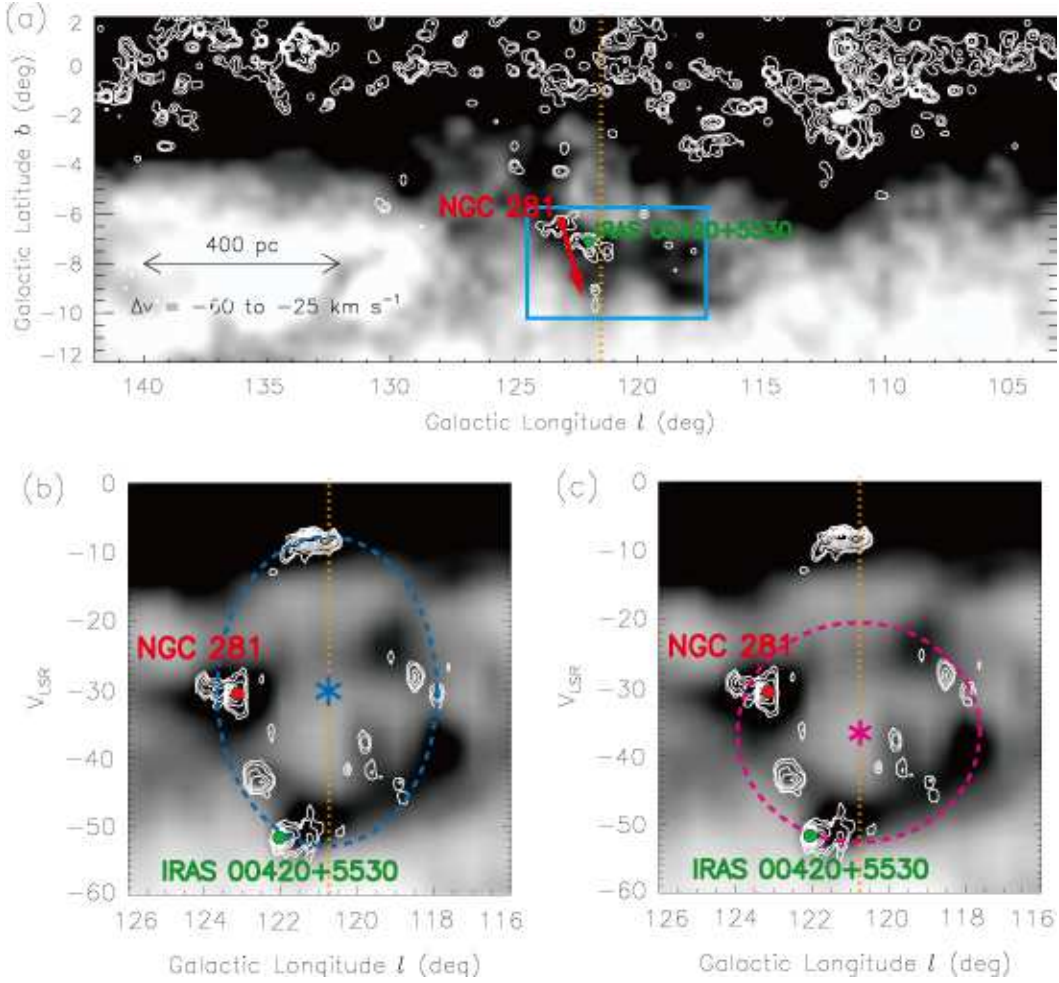}
  \end{center}
  \caption{The NGC~281 superbubble on the reproduced diagrams originally by Megeath \etal\ (2003) (inverse grayscale H\emissiontype{I} data from Hartmann \& Burton 1973; white contour $^{12}$CO data from Dame, Hartmann, \& Thaddeus 2001). The positions of NGC~281 (red) and IRAS~00420+5530 (green) are indicated.
(a) The $b$ vs.\ $l$ map of the region, velocity-integrated for the Perseus-arm line-of-sight velocity range of $V_{\rm{LSR}}=-60$ to $-25$~km~s$^{-1}$.  The measured systemic motion of NGC~281 relative to the rotating Galactic disk is plotted as a red arrow.
(b) (c) The $V_{\rm{LSR}}$ vs.\ $l$ diagram, latitude-integrated for the galactic latitude range indicated by a cyan rectangle in (a). Two possible rings of clouds with their centers are indicated in cyan and pink (see text).}\label{fig:hilv}
\end{figure*}

Using our new results from parallax and proper-motion measurements of the H$_2$O maser source in NGC~281~West, we now revise the motion of the NGC~281 region with respect to the Galactic rotation as reported in Sato \etal\ (2007).

In an analogous manner to Sato \etal\ (2007), we show the modified systemic motion of the NGC~281 region away from the Galactic plane, traced by our H$_2$O maser observations, as a red arrow in figure~\ref{fig:hilv}a, relative to a frame rotating with the Galactic rotation.
Figures~\ref{fig:hilv}a, b and c are an image of the NGC~281 superbubble plotted as $b$ vs.\ $l$ (velocity-integrated for the Perseus-arm line-of-sight velocity range, $V_{\rm{LSR}}=-60$ to $-25$~km~s$^{-1}$) and $V_{\rm{LSR}}$ vs.\ $l$ (latitude-integrated for the galactic latitude range indicated by a cyan rectangle in figure~\ref{fig:hilv}a) diagrams after Megeath \etal\ (2003) and based on H\emissiontype{I} 21-cm line data by Hartmann and Burton (1973) and $^{12}$CO($J=1-0$) line data by Dame, Hartmann, and Thaddeus (2001).
At the position of NGC~281 West, the Galactic plane lies almost parallel to the east-west (RA) direction with a position angle of 90.$^\circ$2,
 so that we can regard the motions in $X$ (RA) and $Y$ (Dec) directions as those parallel ($l$) and perpendicular ($b$; with only small projection effect at $b=-$6.$^\circ$31) to the Galactic plane, respectively.

Subtracting the expected apparent motions of NGC~281 due to the (simulated) Galactic rotation (with respect to the LSR) and the solar motion effect, ($\mu_{{\rm G}X}$,~$\mu_{{\rm G}Y}$)$=$($-$3.27,~$-$0.29) mas~yr$^{-1}$ (including the apparent motion arising from the nonzero galactic latitude of NGC~281; $b=-$6.$^\circ$31) and ($\mu_{{\rm S}X}$,~$\mu_{{\rm S}Y}$)$=$(0.84,~$-$0.53) mas~yr$^{-1}$, respectively, from the measured mean proper motion of the H$_2$O maser source in NGC~281~West yields the systemic motion of the region with respect to the rotation of the Galaxy.
Here we have adopted our new distance estimate of NGC~281, 2.82~kpc, and the solar motion relative to the LSR based on {\it Hipparcos} data (Dehnen \& Binney 1998), and also assumed the distance from the Sun to the Galactic center, $R_0$, to be 8.0~kpc (Reid 1993), the rotational velocity of the Galaxy at the Local Standard of Rest (LSR, at $R_0$), $\Theta_0$, to be 236~km~s$^{-1}$ (Reid \& Brunthaler 2004), and a flat rotation curve of the Galaxy (i.e.\ a differential rotation with a nearly constant rotational velocity $\Theta\approx\Theta_0$ for the outer Galaxy).
Subtracting these two effects of the Galactic rotation and the solar motion on the apparent motion of NGC~281 from the observed proper motions, we obtain the modified mean absolute proper motions of the maser features in C3 and C1 with respect to the Galactic rotation to be: ($\mu_X$,~$\mu_Y$)$_{{\rm C3}-{\rm GR}}=$(0.26,~$-$0.62) mas~yr$^{-1}$ and ($\mu_X$,~$\mu_Y$)$_{{\rm C1}-{\rm GR}}=$($-$0.98,~$-$2.02) mas~yr$^{-1}$, respectively.
The mean of these two spatial components is expected to trace the systemic motion of the NGC~281 region and obtained to be: ($\mu_X$,~$\mu_Y$)$_{{\rm sys}-{\rm GR}}=$($-$0.36,~$-$1.32)~mas~yr$^{-1}$
The direction of this systemic motion with respect to a frame rotating with the Galactic rotation is shown as a red arrow in figure~\ref{fig:hilv}a.
At a distance of 2.82~kpc, a proper motion of 1~mas~yr$^{-1}$ corresponds to a transverse velocity of 13.4~km~s$^{-1}$, and therefore our velocity estimate for the systemic motion of NGC~281 is: ($v_l$,~$v_b$)$=$($-$4.8,~$-$17.7)$\simeq$($-$5,~$-$18)~km~s$^{-1}$ in the directions toward increasing $l$ and $b$, respectively (the minus signs here thus indicate the motions toward decreasing $l$ and $b$).

As mentioned in Sato \etal\ (2007), if we adopt the IAU standard values for the Galactic rotation of $R_0=$8.5~kpc and $\Theta_0=$220~km~s$^{-1}$ (Kerr \& Lynden-Bell 1986) instead of $R_0=$8.0~kpc and $\Theta_0=$236~km~s$^{-1}$, and for the solar motion relative to the LSR of the velocity $V=$19.5~km~s$^{-1}$ toward $\alpha_{\rm 2000}=$271.$^\circ$0, $\delta_{2000}=$29.$^\circ$0, then the resulting values vary by up to $\sim$10$\%$ in the $b$ direction: ($\mu_{{\rm G}X}'$,~$\mu_{{\rm G}Y}'$)$=$($-$2.92~$-$0.25) mas~yr$^{-1}$, ($\mu_{{\rm S}X}'$,~$\mu_{{\rm S}Y}'$)$=$(1.25,~$-$0.61) mas~yr$^{-1}$ and ($\mu'_X$,~$\mu'_Y$)$_{{\rm sys}-{\rm GR}}=$($-$1.12,~$-$1.28)~mas~yr$^{-1}$, yielding a systemic velocity estimate of ($v'_l$,~$v'_b$)$\simeq$($-$15,~$-$17)~km~s$^{-1}$.

In the $l$ direction, we did not discuss in Sato \etal\ (2007) the velocity component $v_l$ parallel to the Galactic plane due to its larger dependence of this direction on the Galactic rotation model as can be seen in the two cases above.
The velocity component $v_l$ toward decreasing galactic longitude obtained above differs from what one would expect from a simple expansion of the NGC~281 superbubble if the origin (e.g., supernova explosions) of the superbubble, expected to be around $l\simeq 121^\circ$ from the ring center in the $V_{\rm{LSR}}$ vs.\ $l$ diagram of figures~\ref{fig:hilv}b and ~\ref{fig:hilv}c (see next section), was rotating with the Galaxy.
The Galactic rotation velocity that would yield $v_l\ge0$ (toward increasing galactic longitude) is $\Theta\ge262$~km~s$^{-1}$ for $R_0=8.0$~kpc; thus the negative velocity component $v_l$ could be explained by a faster rotational velocity of the Galaxy.
However, it is likely that this motion $v_l$ of the region relative to the Galactic rotation is real, reflecting the motion of the superbubble as a whole in a large-scale phenomenon such as velocity jumps over a spiral shock in the Perseus arm of the Galaxy (e.g., Xu \etal\ 2006; see $\S4.3$).

In the $b$ direction, we find the NGC~281 region at a distance of $z\simeq 311$~pc (at $d=2.82$~kpc) from the Galactic plane and moving away from the Galactic plane with a velocity component of $\sim 18$~km~s$^{-1}$.
Assuming the origin of the NGC~281 superbubble to lie within the Galactic plane (at $b=0^\circ$), the dynamical timescale of the superbubble estimated from a simple ballistic motion of $v_b$ is: $t_b=z/v_b=b/\mu_{Y-{\rm GR}}=(-6.^\circ31)/(-1.36)$~mas~yr$^{-1}=16.7\simeq 17$~Myr, where $\mu_{Y-{\rm GR}}$ is the derived $Y$ motion of the maser source in NGC~281 West relative to the Galactic rotation.
This timescale, 17~Myr, is consistent with our timescale estimate of the superbubble expansion in the direction (almost) parallel to the Galactic plane to be discussed in the next section.

Here we also simply estimate the kinetic energy of the NGC~281 region in the direction perpendicular to the Galactic plane, and in the next section we will estimate the expansion velocity, timescale, and energy parallel to the plane as well. 
According to the mass estimates by Megeath \etal\ (2002) of the clouds in the region (indicated by a cyan rectangle in figure~\ref{fig:hilv}a), the velocity-integrated mass in atomic and molecular gas are $3.5\times10^5 M_\odot$ and $10^5 M_\odot$, respectively.
Megeath \etal\ (2002) estimated the total kinetic energy of $4.5\times10^{51}$~ergs for the ring of clouds indicated by a cyan (dashed) ellipse in the ($l$,~$V_{\rm{LSR}}$) diagram of figure~\ref{fig:hilv}b, expanding at a velocity of $v_\mathrm{ring}=$22~km~s$^{-1}$.
Using the same mass estimate as given by Megeath \etal\ (2002) and our new estimate of the velocity component $v_b\sim 18$~km~s$^{-1}$ away from the Galactic plane, we estimate the region to also have kinetic energy of $3.0\times10^{51}$~ergs in the direction perpendicular to the plane, which requires the energy of multiple supernovae.

\subsection{NGC~281 Superbubble: 3D Structure and Expansion}
\begin{figure*}[t]
  \begin{center}
    \FigureFile(130mm,120mm){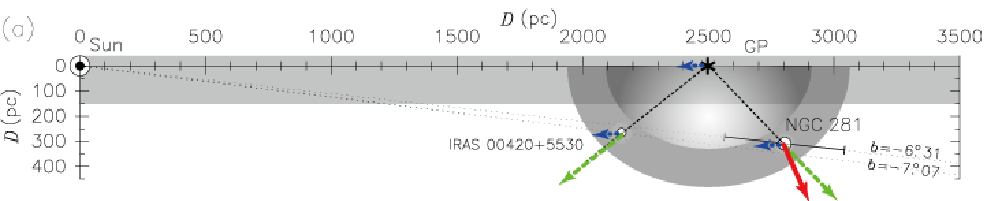}
    \FigureFile(110mm,110mm){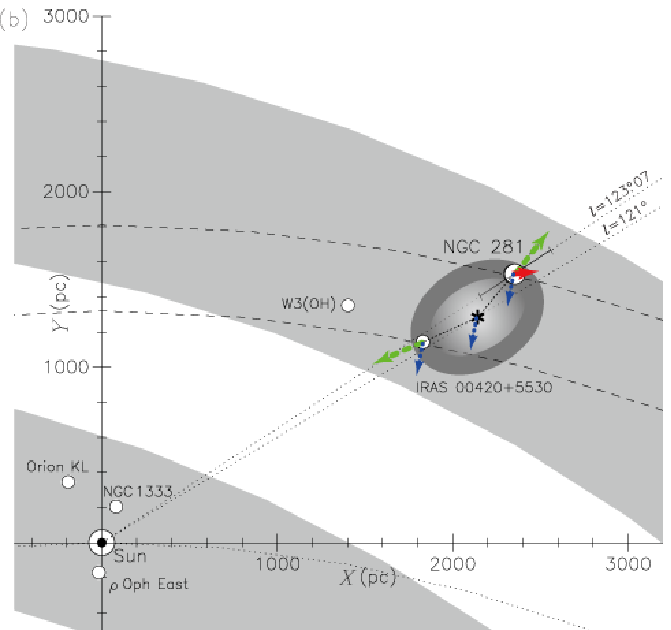}
  \end{center}
  \caption{Schematic diagram of 3D structure and motion of the NGC~281 superbubble: (a) edge-on view (at a galactic longitude of $l\sim 123^\circ$) and (b) face-on view of the Galactic disk. Note that all motion vectors are as seen from the rotating Galactic disk (i.e.\ the Galactic rotation is subtracted). 
The expected motions are plotted as arrows for the explosion origin (blue dotted; assumed to be in the disk), expansion of the superbubble (green dashed) and for the observed motion of NGC~281 (red full).
Errors are also indicated for our distance estimate of NGC~281 (see text).
}\label{fig:3d}
\end{figure*}

Our determination of the parallactic distance to NGC~281 enables us to discuss the 3D structure and expansion of the NGC~281 superbubble.

In the NGC~281 superbubble, there exists another H$_2$O maser souce in the star-forming region IRAS~00420+5530 ($\alpha_{2000}=$00$^{\rm h}$45$^{\rm m}$, $\delta_{2000}=+$55$^\circ$\timeform{47'} or $l=$122$^\circ$.01, $b=-7.^\circ07$), whose position is indicated by green dots in figure~\ref{fig:hilv}.
Moellenbrock, Claussen, and Goss (2007) measured the parallactic distance of the H$_2$O maser source in IRAS~00420+5530 with the Very Long Baseline Array (VLBA) to be $d=2.17\pm0.05$~kpc.
Combining the two parallax results, one can examine the size, structure and the expansion of the superbubble three-dimensionally.
Figures~\ref{fig:3d} illustrates the 3D positions of the two sources, NGC~281 and IRAS~00420+5530, in (a) edge-on (as a cross section at a galactic longitude $l\sim 123^\circ$) and (b) face-on (as seen from the north Galactic pole) views of the Galaxy.

As mentioned previously, Megeath \etal\ (2002, 2003) found an expanding ring of molecular clouds (called {`}Megeath's ring{'} hereafter), and estimated from $\Delta l$ (i.e.\ the ring size in galactic longitudes $l$) to be 270~pc.
From $\Delta V_{\rm{LSR}}$ (i.e.\ the ring {`}size{'} in line-of-sight velocities $V_{\rm{LSR}}$) of $v_\mathrm{ring1}\simeq$22~km~s$^{-1}$ and from $\Delta l/\Delta V_{\rm{LSR}}$ the dynamical age is $\sim6$~Myr.
However, since the heliocentric distances to NGC~281 and IRAS~00420+5530 are now determined by the parallax measurements to be 2.82~kpc and 2.17~kpc, respectively, these clouds (shown as red and green points in figure~\ref{fig:hilv}) are separated from each by 650~pc ($\pm$290~pc) along the line of sight.
This then implies the major axis of Megeath's ring is along the line of sight and a size of approximately $700-1900$~pc in diameter (as derived geometrically from figure~\ref{fig:hilv}b; the radius of Megeath's ring can be approximated by the distance between NGC~281 and IRAS~00420+5530).
However, it remains unclear whether such an elongated shape of the molecular ring is plausible, compared to the apparent ring size of $\sim 300$~pc (in part of the NGC~281 superbubble with apparent size of $300-500$~pc) as seen in the ($l$,~$b$) plane (figure~\ref{fig:hilv}a).

We thus suggest a smaller ring of clouds indicated by a pink (dashed) ellipse (called {`}our ring{'} hereafter) in figure~\ref{fig:hilv}c, excluding a cloud at $V_{\rm{LSR}}\lesssim -10$~km~s$^{-1}$ from Megeath's ring, as another possible interpretation of the ($l$,~$V_{\rm{LSR}}$) diagram.
In this interpretation, NGC~281 and IRAS~00420+5530 are close to the far side and near side of the expanding ring, yielding a smaller ring diameter of $\sim 650$~pc along the line of sight, a lower ($\sim$70\%) expansion velocity of the ring, $v_\mathrm{ring2}\simeq 15$~km~s$^{-1}$, and thus a lower ($\sim$50\%) kinetic energy of $2.1\times10^{51}$~ergs than for Megeath's ring.
The sum of this kinetic energy of the ring expansion (almost) parallel to the Galactic plane and the kinetic energy perpendicular to the plane (estimated to be $3.0\times10^{51}$~ergs in the previous section) yields total kinetic energy of $5.1\times10^{51}$~ergs, again requiring multiple supernovae.
We also estimate the timescale of the ring expansion parallel to the plane, $t_\mathrm{ring2}\simeq 325$~pc$/v_\mathrm{ring2}\simeq 21$~Myr, which is in good agreement with the timescale $t_b\simeq 17$~Myr independently estimated in the previous section from the velocity component $v_b$ perpendicular to the Galactic plane.

Note that in either interpretation of these two rings, the center of the ring (denoted by $*$ in figures~\ref{fig:hilv}b and \ref{fig:hilv}c) is at the same galactic longitude, $l\simeq 121^\circ$ (as indicated by yellow dotted lines in figure~\ref{fig:hilv}).
Considering the ring center as the origin of the superbubble, we can compare the (observed) line-of-sight velocities of the origin (i.e.\ the ring center in figures~\ref{fig:hilv}b and \ref{fig:hilv}c), NGC~281 and IRAS~00420+5530 with those expected from the Galactic rotation model, expressed by 

\begin{equation}\label{eq:Vlsr1}
\displaystyle V_{\scriptsize \mbox{LSR}}=R_0 \sin l \cos b \left(\frac{\Theta}{R}-\frac{\Theta_0}{R_0} \right)+V_z\sin b,
\end{equation}
where $R$ and $\Theta$ are the galactocentric distance and rotational velocity of the Galaxy at the source position ($l$,~$b$) in the galactic coordinates, and those for the LSR are $R_0$ and $\Theta_0$.
Here we also included $V_z$, the source velocity component perpendicular to the Galactic plane (toward the north Galactic pole), in order to take into account the projection effects on the line-of-sight velocities for sources at non-zero galactic latitudes.
However, we neglect this projection-effect term because $\sin b\simeq -0.1$ for NGC~281 and IRAS~00420+5530 and also because our measurement gives $V_z\simeq v_b\simeq -18$~km~s$^{-1}$ of NGC~281 smaller than the galactic rotation ($\Theta$ and $\Theta_0$) by an order of magnitude.
Using the heliocentric distance to the source, $d$, its projection onto the Galactic plane, $D'=d\cos b$, and the cosine formula, $R$ is expressed by  
\begin{eqnarray}
\displaystyle R&=&\sqrt{D'^2+R_0^2-2D'R_0\cos l}\\
&=&\sqrt{d^2 \cos^2 b+R_0^2-2R_0 d\cos l\cos b}.
\end{eqnarray}
Again we adopt $\Theta\simeq\Theta_0\simeq 236$~km~s$^{-1}$, $R_0\simeq 8.0$~kpc, and then the line-of-sight velocities expected from the Galactic rotation model are: $V'_{\rm{N281}}=-36$~km~s$^{-1}$ for NGC~281, $V'_{\rm{I00420}}=-28$~km~s$^{-1}$ for IRAS~00420+5530, and $V'_{\rm{ring1}}=-36$~km~s$^{-1}$ for Megeath's ring center (expected to be at the same heliocentric distance $d=2.82$~kpc as NGC~281 from figure~\ref{fig:hilv}b), and $V'_{\rm{ring2}}=-33$~km~s$^{-1}$ for our ring center, assumed to be at $d=2.5$~kpc.
The observed line-of-sight velocities are: $V_{\rm{LSR}, N281}=-31$~km~s$^{-1}$ for NGC~281 (in $^{12}$CO lines; Lee \& Jung 2003), which is redshifted from the model by $\Delta V_{\mathrm{LSR, N281}}\simeq +5$~km~s$^{-1}$, and for IRAS~00420+5530, on the other hand, $V_{\rm{LSR}, I00420}=-51$~km~s$^{-1}$ (in C$^{34}$S(3$-$2) and other lines; Brand\ et al.\ 2001), which is blueshifted from the model by $\Delta V_{\mathrm{LSR, I00420}}\simeq -23$~km~s$^{-1}$.
This large blueshift of IRAS~00420+5530 is likely due to the expansion of the molecular ring in the superbubble.
The redshift of NGC~281 is capable of two different interpretations using the Megeath's ring model and ours.

For Megeath's ring, the line-of-sight velocity of the ring center (in figure~\ref{fig:hilv}b), $V_{\mathrm{LSR,ring1}}=-30$~km~s$^{-1}$, is also redshifted from the model by $\Delta V_{\mathrm{LSR,ring1}}=+6$~km~s$^{-1}$, suggesting that not only NGC~281 but also the molecular ring as a whole has a redshifted line-of-sight velocity component relative to the Galactic rotation.
Then the ring expansion velocity is $v_{\mathrm{ring1}}\simeq |\Delta V_{\mathrm{LSR, I00420}}-\Delta V_{\mathrm{LSR,ring1}}|\simeq 29$~km~s$^{-1}$.

For our ring, on the other hand, the line-of-sight velocity of the ring center (in figure~\ref{fig:hilv}c), $V_{\mathrm{LSR,ring2}}=-36$~km~s$^{-1}$, is blueshifted from the model velocity by $\Delta V_{\mathrm{LSR, ring2}}\simeq -3$~km~s$^{-1}$, indicating the redshift of NGC~281 is due to the ring expansion.
Then the ring expansion velocity is obtained to be $v_{\mathrm{ring2}}\simeq|\Delta V_{\mathrm{LSR, I00420}}-\Delta V_{\mathrm{LSR,ring2}}|=20$~km~s$^{-1}$, and the timescale of expansion is roughly: $t_\mathrm{ring2}\simeq 325$~pc$/v_\mathrm{ring2}\simeq16$~Myr, in good agreement with our independent timescale estimates above for ring expansion along the Galactic disk ($t_\mathrm{ring2}\sim 21$~Myr) and for vertical expansion out of the disk ($t_b\sim 17$~Myr).

As mentioned in the previous section, the superbubble as a whole seems to have small systemic velocity deviations ($\sim 10$~km~s$^{-1}$) from the Galactic rotation, not only along the line of sight (toward blueshift) but also in the direction toward decreasing galactic longitude.
This motion relative to the Galactic rotation model is plotted as blue dotted arrows in figures~\ref{fig:3d}a and \ref{fig:3d}b with expected expanding motion as green dashed arrows and the resulting observed velocity of NGC~281 as red arrows. 
The deviation of the superbubble motion from the Galactic rotation, in particular the motion toward decreasing galactic longitude (against the expected expanding motion toward increasing longitude), might be reflecting the peculiar motion of the Perseus spiral arm, which may be explained by velocity jumps over a spiral shock in the Perseus arm.
Peculiar motion in the Perseus arm is also reported by Xu \etal\ (2006) for the massive star-forming region W3(OH) (the position indicated in figure~\ref{fig:3d}b) to be rotating slower than the Galactic rotation by $\simeq 14$~km~s$^{-1}$.
For the NGC~281 superbubble, this lends support to the likely peculiar motion of the whole superbubble inward the Galactic rotation and toward the Sun.

Also in figure~\ref{fig:3d}b, we have plotted star-forming regions in the solar neighborhood and in the Perseus arm of the Galaxy with recent parallactic distances precisely measured with VERA and the VLBA: Orion~KL (Hirota \etal\ 2007; Menten \etal\ 2007; Kim \etal\ 2008), NGC~1333 (Hirota \etal\ 2008), IRAS~16293-2422 in $\rho$~Oph~East (Imai \etal\ 2007) and W3(OH) (Xu \etal\ 2006; Hachisuka \etal\ 2006), demonstrating the high capability of the VLBI techniques to reliably trace out the spiral arms of the Galaxy.
Our parallactic distance to NGC~281 is likely to trace the far side of the Perseus arm, and we expect to reveal the detailed structure and motion of the Perseus arm with future parallax measurements with VERA.

The parallactic distances to NGC~281 and IRAS~00420+5530 revealed the size of the superbubble to be above $\sim650$ pc in diameter along the line of sight, almost parallel to the Galactic plane.
Compared to the superbubble size $z$ perpendicular to the Galactic plane, $z\sim300-400$~pc, the superbubble has a structure likely elongated along the plane, or at least spherical (i.e.\ by taking the lower limit of 360~pc in size along the line of sight) and not elongated vertically in the $z$ direction. 
This is contrary to expectation, i.e.\ one would expect a superbubble to expand and blowout in the $z$ direction with a vertical density gradient and with a long timescale of $\sim20$~Myr (see e.g., Tomisaka \& Ikeuchi 1986 and Mac~Low \etal\ 1989 for superbubble models with vertical density stratifaction).

It is thus indicated that the expansion of the NGC~281 superbubble is confined to the Galactic disk.
This may be explained by Tomisaka (1998)'s 3-dimensional MHD simulations which show that the superbubble expansion can be confined to the disk by the magnetic field of the disk, and our estimated 3D structure of the NGC~281 superbubble shows a very analogous shape, size and timescale to those of model~A (size $R_z\sim 255$~pc perpendicular to the plane, timescale $t\sim 22$Myr) of Tomisaka (1998), which assumes a uniform density of the ISM, $n=0.3$~cm$^{-3}$ and a uniform magnetic field, $B(z)=B(0)=5\mu$G, aligned to one direction within and along the disk.
Therefore, the shape of the NGC~281 superbubble sugessts that the superbubble expansion may be confined to the disk due to the magnetic field of the disk.

Within the disk, the Tomisaka (1998) model predicts the elongation of the superbubble to be along the direction of the magnetic field of the disk, which could be in disagreement with the NGC~281 case if the magnetic field lies along the Perseus spiral arm (e.g., Han \etal\ 2006).
The density decrease of the Perseus spiral arm might be also responsible for the structure formation at the position and heliocentric distance of NGC~281, thus yielding a possible implication on the width of the Perseus spiral arm. 
The further details of the 3D structure of the NGC~281 superbubble and its relation to the Perseus arm should be revealed by future parallax measurements with VERA of multiple maser sources in the superbubble.

\subsection{Distance to NGC~281 and the H-R Diagram}
\begin{figure*}[t]
  \begin{center}
    \FigureFile(160mm,150mm){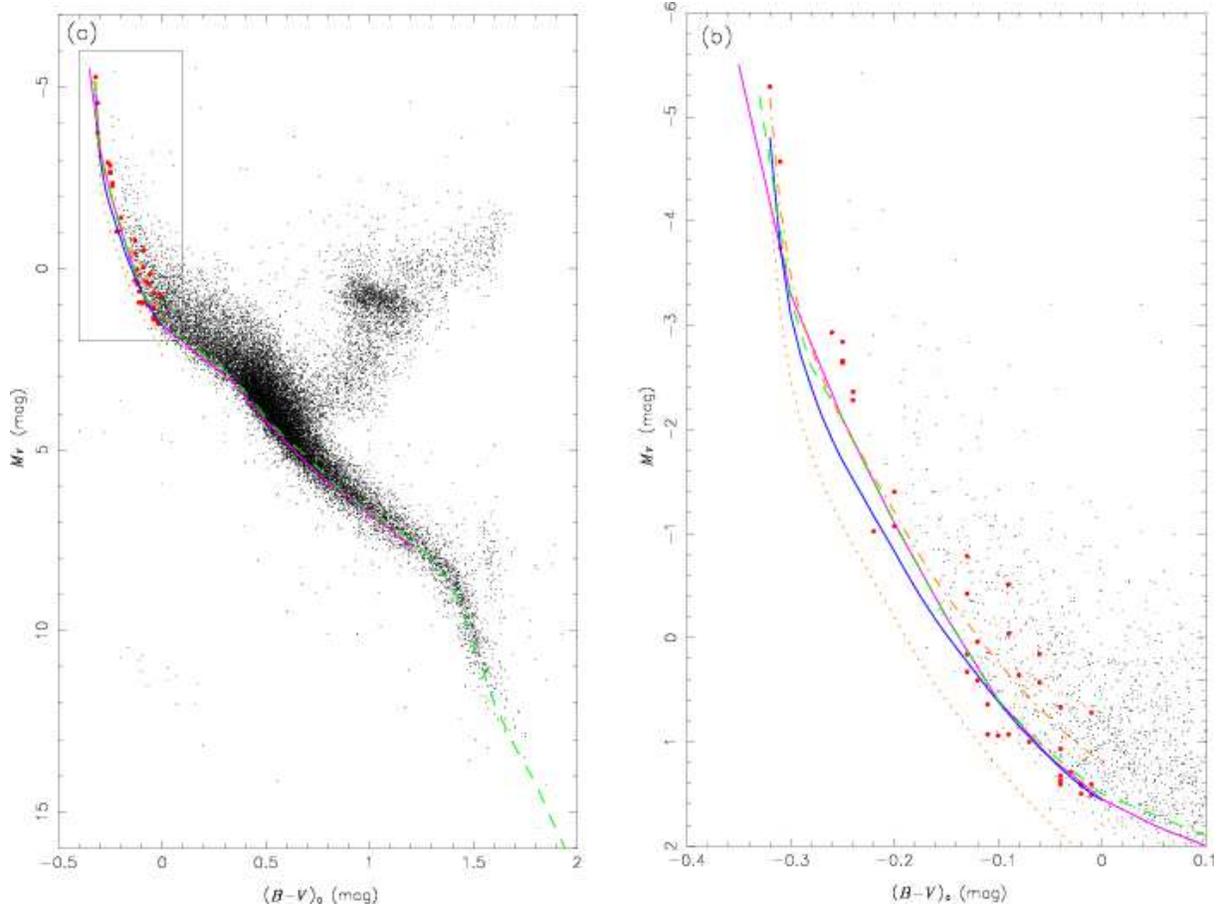}
  \end{center}
  \caption{(a) The H-R (color-magnitude) diagram of stars with parallactic distances determined with precisions better than 10\% (i.e.\ $\sigma_\pi < 0.10\pi$).  Black dots show the {\it Hipparcos} data (Perryman \etal\ 1997), and red dots show the photometric data of ZAMS of early-type stars in NGC~281 by Guetter \& Turner (1997), plotted with absolute magnitudes derived with our new parallactic distance to NGC~281 of 2.82~kpc.
Empirical ZAMS curves are also plotted: Turner (1976) in blue; Blaauw (1963) in pink; and Schmidt-Kaler (1982) in green (dashed).
Orange dotted (lower) and dashed (upper) curves show ZAMS curves that would yield distances of NGC~281 to be 2.2~kpc and 3.5~kpc (see text).
(b) Magnified view of the boxed area (i.e.\ the high-mass end) in (a).}\label{fig:HR}
\end{figure*}

Our astrometric determination of the distance to NGC~281 also provides important information toward better understanding of early-type (O$-$B) high-mass ($M\gtrsim10\Mo$) stars and stellar evolution in combination with photometric studies of the OB stars in NGC~281 (e.g., Henning \etal\ 1994; Guetter \& Turner 1997).

In the 1990s, the {\it Hipparcos} astrometry satellite measured the trigonometric parallaxes of more than 100~000 stars in the solar neighborhood with a precision of about 1~mas, including direct distance estimates of 20~853 stars with a precision better than 10 percent and of 49~399 stars (in total) with a precision better than 20 percent (Perryman \etal\ 1995, 1997).
Figure~\ref{fig:HR} is the H-R (Hertzsprung-Russell) diagram or the color-absolute magnitude diagram based on the {\it Hipparcos} data of stars (shown as black dots) with parallactic distances determined with precisions better than 10\%.

The achievements of {\it Hipparcos} marked a significant milestone in the history of astronomy and brought a new era of astrometry, since before {\it Hipparcos}, parallax measurements were limited to the nearest stars and thus direct distance estimates of only several hundred nearby stars were known with precisions better than 20 percent. 
Nevertheless, the parallax measurements by {\it Hipparcos} were restricted only to within a few hundred parsecs from the Sun, which is much smaller than the size of the Galaxy (e.g., $\sim$15~kpc in disk radius).

Due to the low local space density of OB stars, there is a lack of early-type high-mass stars with absolute magnitudes reliably derived from astrometrically measured distances, as seen on the left side (i.e.\ the high-mass end) of figure~\ref{fig:HR}. 
Alternatively, most of the distances and thus absolute magnitudes of OB stars are estimated by a comparison of photometric colors and magnitudes with an empirical standard of zero-age main sequence (ZAMS) stars in the color-magnitude diagram, derived from other well-studied objects such as Hyades (see Anthony-Twarog 1982).
Such ZAMS comparisons assume compositional differences in clusters can be ignored, however, the metalicity dependence of ZAMS stars is important when used as distance scale and is the cause of difficulty in a direct comparison between theoretical models and empirical ZAMS curves (e.g., Anthony-Twarog 1982; Zinnecker \& Yorke 2007).
Therefore, precise and direct determination of distances, and thus absolute magnitudes, are desired and essential for further studies of stellar evolution sensitivity to metalicity.

As mentioned earlier, extremely high astrometric accuracy can now be achieved with VLBI techniques out to as kiloparsec distances with precisions better than 10 percent (e.g., Xu \etal\ 2006; Hachisuka \etal\ 2006; Menten \etal\ 2007).
In particular, recent results of parallax measurements with VERA are reported in Hirota \etal\ (2007, 2008a, 2008b), Honma \etal\ (2007), Nakagawa \etal\ (2008), Kim \etal\ (2008), and Choi \etal\ (2008), clearly demonstrating VERA's high capability for the Galaxy-scale astrometry.

Using our new distance determination of NGC~281 by parallax measurements with VERA, $d=2.82\pm0.24$~kpc, to derive absolute magnitudes, we have plotted the photometric data of ZAMS of early-type stars in NGC~281 by Guetter \& Turner (1997) in figure~\ref{fig:HR} as red dots, which give improved coverage of the high-mass end of the H-R diagram.
In figure~\ref{fig:HR}, the empirical ZAMS curves are also plotted for reference: Turner (1976) in blue; Blaauw (1963) in pink; and Schmidt-Kaler (1982) in green (dashed).
Orange dotted (lower) and dashed (upper) curves in figure~\ref{fig:HR} show ZAMS curves that would yield distances of NGC~281 to be 2.2~kpc and 3.5~kpc (i.e.\ the distance discrepancy of NGC~281 from previous studies), made by additionally shifting the ZAMS by Turner (1976), from which Guetter \& Turner (1997) obtained a photometric distance to NGC~281 of 2.94~kpc.

Figure~\ref{fig:HR} clearly shows the NGC~281 data agree well with the empirical ZAMS curves, and that resolving the large distance discrepancy of NGC~281 from previous studies was important.
The agreement of the parallactic distance with the photometric distance is important for confirming the accuracy of the correction for extinction or reddening in photometric studies such as Guetter \& Turner (1997) and for enhancing the precision of the H-R diagram.

Our determination of the parallactic distance to NGC~281 therefore contributes to the extension of the H-R diagram for the study of high-mass stars and, furthermore, opens up new possibilities of using the H-R diagram with parallax measurements for Galaxy-scale studies, for example, investigating the formation of the Galaxy via the difference in the H-R diagram for components such as the bulge and spiral arms.  
Further contributions are expected with the VLBI techniques and with VERA for direct parallax determination far beyond the solar neighborhood toward a comprehensive understanding of the Galaxy.  

\section{Conclusions}
In this paper, we have presented the results of our multi-epoch phase-referencing observations with VERA over 18 months of an H$_2$O maser source in the Galactic star-forming region NGC~281~West, associated with a fragmenting superbubble $\sim300$~pc above the Perseus spiral arm.
The primary results are summarized as follows:

\begin{enumerate}
\item We detected a trigonometric parallax of 0.355$\pm$0.030~mas, corresponding to a distance of 2.82$\pm$0.24~kpc to NGC~281.
Our parallactic distance agrees well with the photometric distance of 2.94$\pm$0.15~kpc derived by Guetter \& Turner (1997), allowing for improved study of the absolute magnitutudes of high-mass OB stars, and resolved the large distance discrepancy of NGC~281 from previous photometric and kinematic studies.\\

\item We revised the absolute proper motions of the H$_2$O maser features measured in Sato \etal\ (2007), and using our new determination of the parallactic distance, we derived more precisely the velocity component of the NGC~281 region $v_b\simeq 18$~km~s$^{-1}$ perpendicular to the Galactic plane.
This yields a timescale of $t_b\sim 17$~Myr and kinectic energy of the region to be $3.0\times10^{51}$~ergs in the direction perpendicular to the plane.\\

\item
We demonstrated the 3D structure and expansion of the NGC~281 superbubbles in comparison of our parallactic distance to NGC~281 with the parallactic distance derived by Moellenbrock \etal\ (2007) for another H$_2$O maser source, IRAS~00420+5530 in the superbubble.
Our new parallactic distance revealed the structure of the superbubble $\sim$650 pc in size parallel to the Galactic disk and with a shape slightly elongated along the disk or spherical, but not vertically elongated in the $z$ direction. 
Therefore, the superbubble expansion may be confined to the disk, possibly due to the magnetic field of the disk. \\

\item
We suggested a new possible interpretation of an expanding molecular ring parallel to the Galactic plane in the longitude-velocity diagram by Megeath \etal\ (2002, 2003).
In either interpretation, the ring center, i.e.\ the likely origin of the superbubble lies at $l\simeq 121^\circ$.
The velocity deviation of the superbubble from the Galactic rotation is estimated to be $\sim 10$~km~s$^{-1}$ inward the Galactic rotation and toward the Sun from our proper motion measurements and from the longitude-velocity diagram, which might be due to velocity jumps over a spiral shock in the Perseus spiral arm.\\

\item
We also estimated the velocity component and timescale of the ring expansion to be $v_\mathrm{ring2}\simeq 15-20$~km~s$^{-1}$ and $t_\mathrm{ring2}\sim 16-21$~Myr, respectively, parallel to the Galactic plane, which are in good agreement with those independently estimated for the direction perpendicular to the Galactic plane, $v_b\simeq 18$~km~s$^{-1}$ and $t_b\sim 17$~Myr.
The kinetic energy estimate of the region parallel to the plane is $2.1\times10^{51}$~ergs, and the total kinetic energy of both perpendicular and parallel to the plane is then estimated to be $5.1\times10^{51}$~ergs, requiring multiple supernovae.
\end{enumerate}

\medskip
\noindent
We are deeply grateful to the referee Dr.~Felix~J.~Lockman for his invaluable comments and suggestions that improved both scientific discussions and presentation of the manuscript.
We would like to thank Dr.~Mark~J.~Reid for his tremendous help with carefully reading and improving the manuscript and for his insightful comments on scientific issues.
We are very grateful to Prof.~Kohji~Tomisaka for illuminating and insightful discussions.
We also greatly appreciate the optical image for figure~1 that was kindly provided by Dr.~Travis~A.~Rector and the National Optical Astronomy Observatory (NOAO). 
We wish to thank all the support staff at VERA for their efforts and continuous support for our observations. 
M.~Sato gratefully acknowledges financial support by a Research Fellowship from the Japan Society for the Promotion of Science (JSPS) for Young Scientists.
This research has made use of the SIMBAD database, operated at CDS, Strasbourg, France and of NASA's Astrophysics Data System.

\end{document}